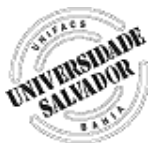

# UNIVERSIDADE SALVADOR – UNIFACS
## NÚCLEO DE PESQUISA INTERDEPARTAMENTAL EM REDES DE COMPUTADORES (NUPERC)
### MESTRADO EM REDES DE COMPUTADORES

**MARCOS PORTNOI**

# CRIPTOGRAFIA COM CURVAS ELÍPTICAS

Salvador – BA
2005

# CRITOGRAFIA COM CURVAS ELÍPTICAS[*]

Marcos Portnoi[**]

Orientador: Prof. Rafael T. de Souza Jr.[***]

**Resumo**

**Abstract**

Este trabalho apresenta o uso das curvas elípticas em criptografia. Sua segurança está baseada no problema do logaritmo discreto. Este problema aparentemente é significativamente mais difícil de resolver, comparado com o problema do logaritmo discreto usado por outros sistemas de criptografia. É dada uma visão geral de sistemas de criptografia comuns, como Diffie-Hellman e RSA, e discute-se um esquema de criptografia usando curvas elípticas.

This paper presents an overview of the use of elliptic curves in cryptography. The security of this cryptosystem is based on the discrete logarithm problem, which appears to be much harder compared to the discrete logarithm problem in other cryptosystems. An overview of common cryptosystems is given, such as Diffie-Hellman and RSA, and an elliptic curve cryptography scheme is discussed.

**Palavras-Chave**: Curvas elípticas, criptografia, sistema de criptografia, problema logaritmo discreto, Diffie-Hellman, RSA.

---

## Introdução

Criptografia é a ciência que trata de cifrar a escrita, de modo a torná-la ininteligível para os que não tenham os métodos convencionados para ter acesso a ela. Em Tecnologia da Informação, esta definição é ampliada a fim de englobar não só a escrita, mas qualquer tipo de documento ou dados que devam ser tratados secretamente. Um Sistema de Criptografia define um sistema em que um texto (ou documento, ou qualquer tipo de dado) é transformado através da criptografia em um texto cifrado (*ciphertext*) ou o texto cifrado é transformado de volta à informação original [1], através de algoritmos. A primeira ação é denominada *cifragem* ou *criptografar*, e a segunda é chamada de *decifração* ou *decriptação*.

Os sistemas de criptografia são classificados – [2] *apud* [1] – em *simétricos* e *assimétricos*. Os sistemas simétricos usam uma chave secreta, conhecida por ambos os lados A e B da comunicação e por ninguém mais. Os sistemas assimétricos usam um par de chaves, uma secreta e uma pública, para cada lado A e B da comunicação. A chave secreta de A só é conhecida por este, assim como a chave secreta de B só é conhecida por B. As chaves públicas de A e B são conhecidas por todos.

No sistema simétrico, a comunicação privada entre as duas partes é estabelecida usando uma única chave secreta, que deve ser conhecida somente pelos participantes da comunicação. Esta chave é usada tanto para cifragem como para decriptação do *ciphertext*. Um grande problema desse sistema é que os interlocutores A e B precisam combinar previamente a chave secreta a ser usada através de um meio seguro de comunicação. Se a chave secreta for transmitida por um interlocutor a outro através de um meio inseguro, todo o sistema de segurança estará comprometido. Outra desvantagem é que cada interlocutor deve armazenar uma chave secreta a fim de manter comunicação privada para cada interlocutor diferente, resultando numa grande quantidade de chaves privadas diferentes a armazenar (na ordem de $n^2$ chaves para todo o sistema) [1]. A Figura 1 traz uma ilustração do funcionamento do sistema simétrico.

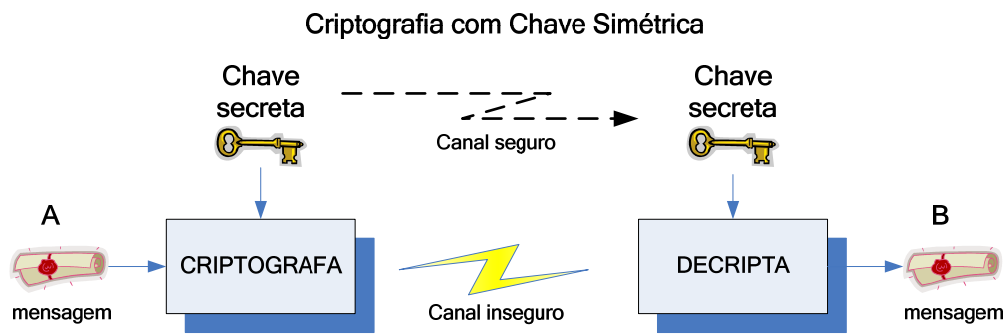

**Figura 1: Sistema de Criptografia com chave simétrica (secreta).**

No sistema de criptografia assimétrico, há duas chaves distintas para cada interlocutor A e B: uma privada, que somente o interlocutor conhece, e uma pública, que pode ser distribuída para qualquer um. Uma chave se presta a cifrar a mensagem e a outra, para decifrá-la. O par de chaves são determinadas por um algoritmo baseado em uma "função de uma via". A função de uma via é definida como uma função $f$ onde, para cada $x$ no domínio de $f$, $f(x)$ é facilmente computável. Porém, para praticamente todo $y$ na imagem de $f$, é computacionalmente inexeqüível[1] encontrar um $x$ que satisfaça $y = f(x)$ [1]. Assim, dado $x$, acha-se facilmente $f(x)$; porém dado $f(x)$, o poder computacional requerido para o cálculo de $x$ é extremamente elevado. O conceito foi introduzido por Diffie e Hellman em 1976 [3], ao proporem o protocolo de troca de chaves que leva seu nome. A segurança do sistema, pois, baseia-se essencialmente na dificuldade de se calcular a função inversa.

No sistema assimétrico, se A deseja enviar uma mensagem secreta para B, então A usa a chave pública de B para criptografar a mensagem. Ao receber a mensagem, B usa sua própria chave privada para decriptá-la. Num sistema alternativo de autenticação, A pode mandar uma mensagem para B, cifrando-a com sua chave secreta (chave secreta de A). Qualquer pessoa, inclusive B, pode decifrar esta mensagem através da chave pública de A (que é, pois, de domínio público). Como somente A detém a chave privada de A, então, se a mensagem puder ser decifrada com a chave pública de A, então a origem foi autenticada (prova-se que somente A pode ter enviado a mensagem).

---

[1] Isto não significa que é computacionalmente impossível de se reverter o cálculo das funções. A segurança destes sistemas de criptografia baseia-se no fato de que a força computacional necessária para a quebra do

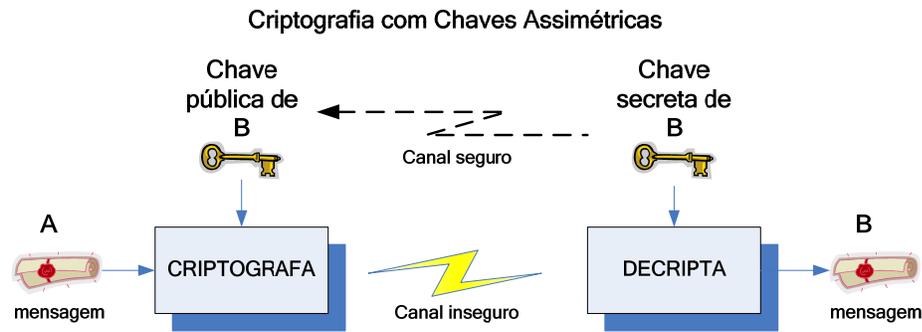

**Figura 2: Sistema de Criptografia com chaves assimétricas (chave pública e chave privada).**

A Figura 2 ilustra o funcionamento do sistema de criptografia com chaves assimétricas, e a Figura 3, uma representação da utilização do sistema para autenticação de remetente. O sistema de criptografia assimétrico resolve o problema de troca prévia de chaves privadas através de um meio inseguro sofrido pelo sistema simétrico. As chaves públicas podem ser livremente comunicadas por meios inseguros, pois somente com as chaves privadas (que só são de conhecimento de seus proprietários) o ciclo de cifração-decifração pode ser completado. Um sistema de criptografia assimétrico largamente usado é o RSA (de Rivest, Shamir e Adleman, seus postulantes), que se baseia na fatoração de números primos grandes [4, 5, 6], e também o sistema conhecido como ElGamal, que usa o problema do logaritmo discreto na implementação de chaves assimétricas, com nível de segurança semelhante do esquema Diffie-Hellman [7].

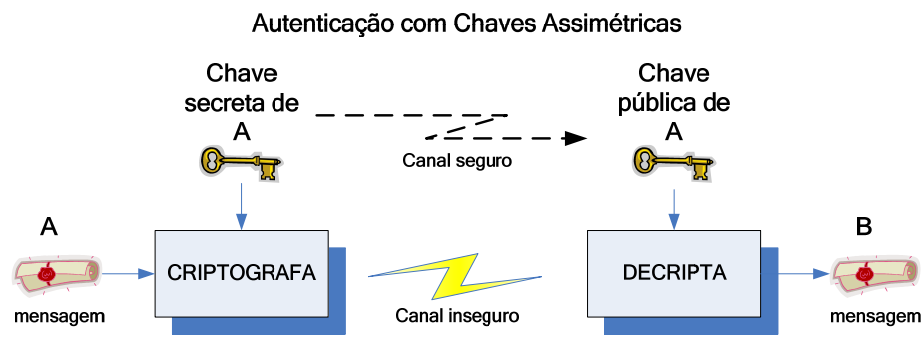

**Figura 3: Sistema de Autenticação usando chaves assimétricas.**

---

sistema é acessível somente para pouquíssimas organizações. Porém, à medida que o poder computacional aumenta graças aos avanços tecnológicos, mais aumenta a possibilidade de quebra da segurança por indivíduos.

Os algoritmos de chaves assimétricas requerem maior tempo computacional que os algoritmos de chave simétrica, mais simples. Uma alternativa é usar um sistema de criptografia simétrica e fazer a troca prévia da chave privada através do protocolo de troca de chaves Diffie-Hellmann, que pode ser realizado por um canal inseguro. A função de uma via proposta por [3] para o protocolo de troca de chaves através de um meio inseguro é $f(x) = \alpha^x \pmod{p}$. Se $y = \alpha^x$, então $x = \log_\alpha(y)$. Encontrar o valor de $x$ nesta equação resulta num problema de cálculo de logaritmo discreto.

---

Seja $p$ um número primo grande e assuma-se que $\alpha$ é um elemento primitivo do campo $Z_p$ (base do logaritmo discreto), onde $p$ e $\alpha$ são públicos.

1. Interlocutor **A** escolhe $x1$ ($0 \leq x1 \leq p - 2$) aleatoriamente.
2. **A** calcula $\alpha^{x1} \bmod p$ e envia o resultado ao interlocutor **B**.
3. **B** escolhe $x2$ ($0 \leq x2 \leq p - 2$) aleatoriamente.
4. **B** calcula $\alpha^{x2} \bmod p$ e envia o resultado ao interlocutor **A**.
5. **A** calcula $K = (\alpha^{x2})^{x1} \bmod p$ e **B** calcula $K = (\alpha^{x1})^{x2} \bmod p$.

**A** e **B** calculam, desta maneira, a mesma chave $K = \alpha^{x1 x2} \bmod p$.

---

**Figura 4: Protocolo de troca de chave Diffie-Hellmann [8].**

## O Sistema RSA: Visão Geral

O algoritmo RSA [5] é de uso constante nos sistemas atuais de criptografia assimétrica. Sua segurança está baseada na dificuldade de se fatorar números inteiros muito grandes e o algoritmo está apresentado na Figura 5.

---

Interlocutor **A** escolhe, secretamente, dois números primos, $p$ e $q$, e publica $n = pq$. Então, **A** aleatoriamente escolhe $b$ tal que $b$ e $\Phi = (p - 1)(q - 1)$ são primos relativos ou primos entre si. **A** calcula $a$ tal que $ab = 1 (\bmod \Phi(n))$. Sua chave secreta é $a$, enquanto que $b$ é revelado publicamente.
Interlocutor **B** cifra sua mensagem $x$ computando:

$$y = x^b \bmod n$$

e envia $y$ para **A**. **A** então decifra e obtém $x$ calculando:

$$x = y^a \bmod n$$

---

**Figura 5: O sistema de criptografia RSA [8].**

Para ser considerado suficientemente seguro, a fatoração de $n = pq$ deve ser computacionalmente inviável. Os números $p$ e $q$ devem ser números primos com tamanho da ordem de 100 dígitos cada. Existem ataques ao RSA que não envolvem algoritmos de fatoração, explorando fraquezas inerentes ao sistema, como escolhas indevidas do número $a$ ou o uso do mesmo número $n$ para comunicação com várias pessoas [8]. Atualmente, considerando o poder computacional acessível ao indivíduo comum, usa-se chaves para o RSA de tamanho 1024 bits.

O sistema baseado em curvas elípticas, tratado a seguir, oferece o mesmo nível de segurança comparativa a um RSA de chave de 1024 bits, usando chaves de tamanho 160 bits. As chaves menores permitem requerimentos menores de memória, menores larguras de banda usadas na transmissão das mensagens, implementação mais rápida e também algoritmos mais velozes. Para a fabricação de chips dedicados a executar o algoritmo de criptografia, as chaves menores resultam em circuitos mais simplificados e portanto menor custo. Em adição, há algoritmos de complexidade de ordem subexponencial para resolução do problema de logaritmo discreto para o RSA. Até o presente, os melhores algoritmos para a resolução do problema de logaritmo discreto para o sistema de curvas elípticas (para curvas escolhidas corretamente) continuam com complexidade de ordem exponencial [6, 9, 10, 11].

## Sistema de Criptografia com Curvas Elípticas

Um sistema de criptografia assimétrica baseado em um grupo de curvas elípticas sobre um campo finito foi primeiramente proposto, de maneira independente, por Koblitz [12 *apud* 1] e Miller [13] em 1985. Concentra-se no problema do logaritmo discreto num grupo formado pelos pontos de uma curva elíptica definida em torno de um corpo de Galois [14]. O melhor algoritmo conhecido para resolução deste problema tem complexidade exponencial, o que confere um alto grau de segurança ao sistema. A definição de uma curva elíptica é a seguinte (os conceitos matemáticos não detalhados neste trabalho poderão ser consultados em [8]):

### Definição de uma Curva Elíptica

Seja $K$ um campo. Por exemplo, $K$ pode ser o campo finito $F_{q'}$ de $F_q$, ou o campo de primos $Z_p$, onde $p$ é um número primo grande, o campo $R$ de números reais, o campo $Q$ de números racionais ou o campo $C$ de números complexos. Uma curva elíptica sobre o campo $K$ é definida pela **equação de Weierstrass**:

$$y^2 + a_1xy + a_3y = x^3 + a_2x^2 + a_4x + a_6 \qquad (1)$$

onde $a_1, a_3, a_2, a_4, a_6 \in K$.

A curva elíptica $E$ sobre $K$ é denominada $E(K)$. O número de pontos em $E$ (a cardinalidade) é denominada $\#E(K)$ ou somente $\#E$.

A equação de Weierstrass pode ser transformada e simplificada para diferentes formas por uma troca linear de variáveis. Para um campo de característica diferente de 2 e de 3, ou seja, para um campo $Z_p$, para $p>2$, a curva elíptica $E$ sobre este campo $Z_p$ é definida por uma equação da forma:

$$y^2 = x^3 + ax + b \qquad (2)$$

onde $a, b \in Z_p$ e $4a^3 + 27b^2 \pmod{p} \neq 0$, de modo que o polinômio não tenha raízes múltiplas, e ainda um elemento **0** chamado **ponto no infinito**.

O conjunto $E(Z_p)$ consiste em todos os pontos $(x,y) \mid x, y \in Z_p$ que satisfazem a equação (1), juntamente com o ponto **0**. A Figura 6 demonstra a aparência de uma curva elíptica.

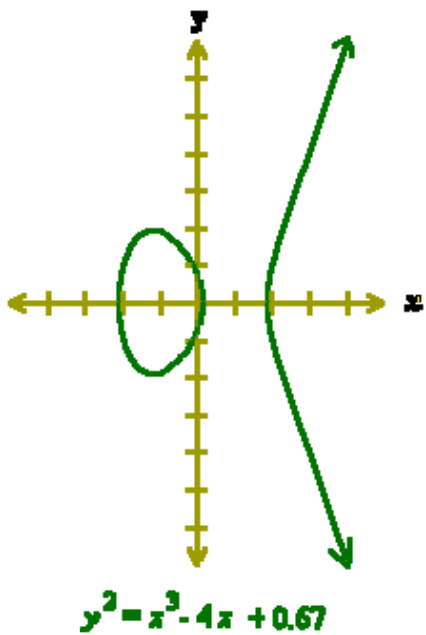

**Figura 6: Exemplo de uma curva elíptica [14].**

Existe uma regra para somar dois pontos pertencentes a uma curva elíptica, de tal forma que esta soma seja um terceiro ponto sobre a mesma curva. O conjunto de pontos *E(Z$_p$)*, juntamente com a operação de soma, formam um grupo abeliano, onde o ponto no infinito **0** é o elemento neutro.

Sejam, pois, P = (*x$_1$ ; y$_1$*) e Q = (*x$_2$ ; y$_2$*) dois pontos distintos tomados em uma curva elíptica *E*. A soma de P e Q, denotada por R = (*x$_3$ ; y$_3$*) é definida através do traçamento de uma linha que atravesse P e Q. Esta linha intercepta a curva elíptica *E* em um terceiro ponto, onde R é a reflexão deste ponto sobre o eixo *x*. Este ponto R é portanto o resultado da operação de soma P + Q.

Se P = (*x$_1$ ; y$_1$*), então o dobro de P, denotado por R = (*x$_3$ ; y$_3$*) define-se pelo traçamento de uma reta tangente à curva elíptica no ponto P. Esta reta intercepta a curva em um segundo ponto, cuja reflexão sobre o eixo *x* é o ponto R [14].

As fórmulas algébricas que representam P + Q podem ser derivadas dos procedimentos geométricos na Figura 7.

---

O inverso de $P = (x_1; y_1) \in E$ é $-P = (x_1; -y_1)$. Se Q ≠ -P, então $P + Q = (x_3; y_3)$, onde

$$x_3 = \lambda^2 - x_1 - x_2$$
$$y_3 = \lambda(x_1 - x_3) - y_1$$

Onde

$$\lambda = \begin{cases} \dfrac{y_2 - y_1}{x_2 - x_1}, \text{ se } P \neq Q \\ \dfrac{3x_1^2 + a}{2y_1}, \text{ se } P = Q \end{cases}$$

---

**Figura 7: Fórmula da adição para curva elíptica [8].**

A Figura 8 ilustra a adição de pontos diferentes e do mesmo ponto numa curva elíptica, conforme definida acima.

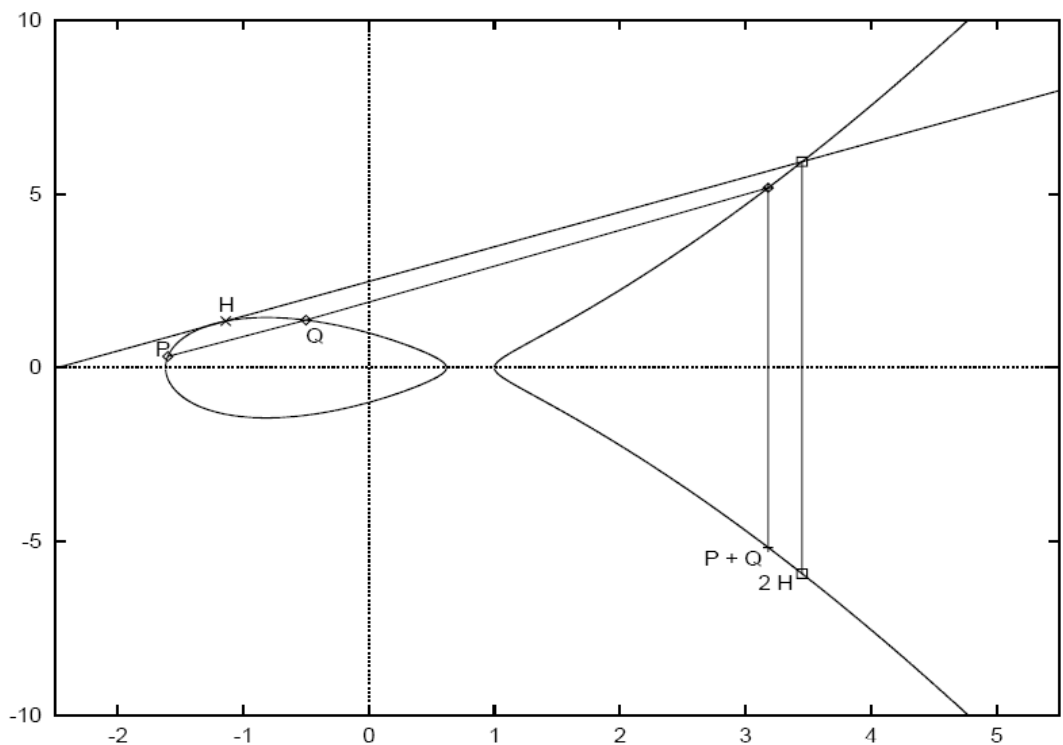

**Figura 8: Adição de dois pontos diferentes P e Q resultando em P + Q e adição do ponto H com ele mesmo, resultando em 2H [9].**

## O Problema do Logaritmo Discreto

Já que uma curva elíptica $E$ é transformada em um grupo abeliano por uma operação de adição (e não por uma operação de multiplicação), a **exponenciação** de um ponto em $E$ refere-se a uma adição repetida. Assim sendo, a enésima potência de $\alpha \in E$ é o enésimo múltiplo de $\alpha$, ou seja, $\beta = \alpha^n = n\alpha$. O **logaritmo** de $\beta$ na base $\alpha$ é o número $n$, o inverso da exponenciação [8].

Para um grupo qualquer $G$, suponha-se que $\alpha, \beta \in G$. Num problema de logaritmo discreto, tenta-se resolver para um inteiro $x$ tal que $\alpha^x = \beta$. Analogamente, no **problema do logaritmo discreto em curvas elípticas**, tenta-se resolver para um inteiro $x$ tal que $x\alpha=\beta$, para $\alpha, \beta \in E$. Para que este problema numa curva sobre $E(F_q)$ seja computacionalmente intratável, é importante selecionar uma curva $E$ e $q$ apropriados tal que $\#E(F_q)$ seja divisível por um número primo grande (de mais de 30 dígitos), ou tal que $q$ seja um número primo grande.

O problema do logaritmo discreto pode ser expresso de outra maneira: dados P e Q, pontos pertencentes ao grupo, encontrar um número *k* tal que *k*P = Q; *k* é denominado o logaritmo discreto de Q na base P. Exemplo: no grupo $Z^*_{13}$, qual é o logaritmo discreto (*x*) de 8 na base 7 [14]. O cálculo é feito iterativamente, conforme ilustrado a seguir.

$7^x$ mod 13 = 8, $x$ = ?

- $7^0$ mod 13 = 1
- $7^1$ mod 13 = 7
- $7^2$ mod 13 = 10
- $7^3$ mod 13 = 5
- $7^4$ mod 13 = 9
- $7^5$ mod 13 = 11
- $7^6$ mod 13 = 12
- $7^7$ mod 13 = 6
- $7^8$ mod 13 = 3
- $7^9$ mod 13 = 8

A solução encontrada iterativamente foi, portanto, o número 9. Para números escolhidos cuidadosamente, a complexidade do algoritmo de descoberta resulta na ordem exponencial, tornando inviável o cálculo de *x*.

**Implementação do Sistema de Criptografia análogo ao ElGamal**

Usando os conceitos de curvas elípticas num sistema de criptografia análogo ao sistema ElGamal [7], uma curva elíptica e um ponto P na curva são escolhidos e tornados públicos. Se um interlocutor **A** deseja se comunicar com **B** (chamar-se-á aqui **A** de Alice e **B** de Bob, de acordo com a convenção quase universalmente adotada na literatura sobre criptografia), então Alice escolhe um inteiro *a* e torna público o ponto *a*P (exponenciação ou adição repetida). Alice mantém o número *a* secreto. Assume-se que a mensagem é composta de pares ordenados de elementos num grupo.

Para transmitir a mensagem ($M_1$ ; $M_2$) para Alice, Bob escolhe um inteiro aleatório *k* e calcula os pontos *k*P e *ak*P = ($x_k$ ; $y_k$). Um *k* diferente deve ser adotado para cada nova mensagem. Então Bob envia para Alice o ponto *k*P e o campo de elementos ($m_1$ ; $m_2$) = ($M_1 x_k$ ; $M_2 y_k$). A mensagem original ($M_1$ ; $M_2$) pode ser decifrada por Alice usando sua chave secreta *a*, através do cálculo de *ak*P a fim de obter os pontos $x_k$ e $y_k$. A mensagem cifrada pode ser decriptada através da divisão, obtendo $M_1 = m_1/x_k$ e $M_2 = m_2/y_k$ [1, 10]. A Figura 9 ilustra todo o processo.

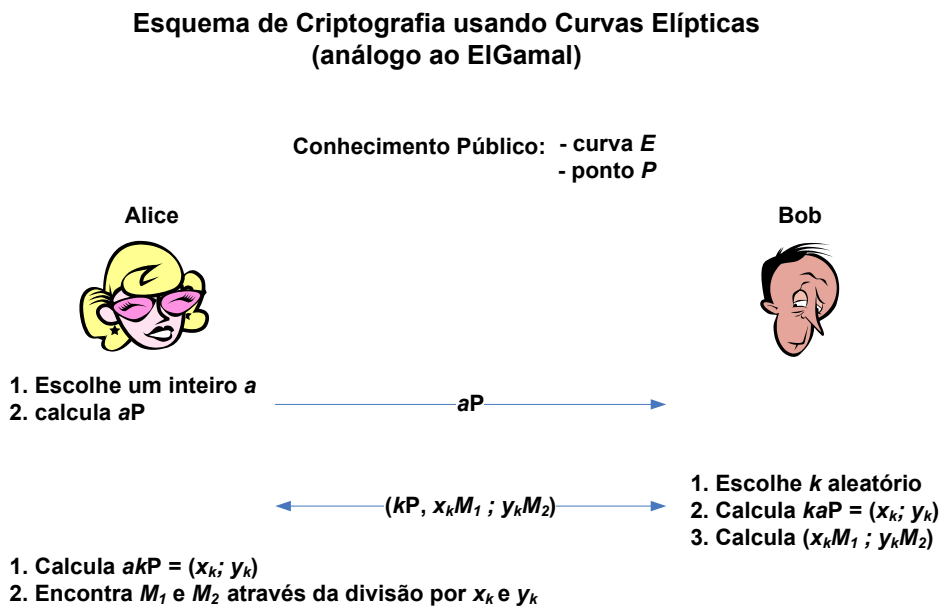

**Figura 9:** Esquema de criptografia baseado em ElGamal e curvas elípticas [1].

## Conclusão

Este artigo descreve brevemente a aplicação das curvas elípticas para um sistema de criptografia de chaves assimétricas. As curvas elípticas provêm uma dificuldade maior para o problema do logaritmo discreto, se comparadas às técnicas comumente usadas de fatoração de números grandes ou o sistema Diffie-Hellman. Isso significa que, para chaves de tamanhos menores, um sistema baseado em curvas elípticas mostra ter um nível de segurança comparável ao sistema RSA com chaves substancialmente maiores.

Apesar da matemática envolvida no conceito de curvas elípticas ser de razoável complexidade, o artigo pretendeu introduzir os conhecimentos mínimos para assimilação do mecanismo de funcionamento do sistema e disponibilizar, no levantamento bibliográfico, referências úteis para maior aprofundamento.

# Referências